\documentclass[times,twocolumn,final]{elsarticle}

\usepackage{amssymb}
\usepackage{amsmath}
\usepackage{cag}
\usepackage{booktabs}
\usepackage{comment}
\usepackage{framed,multirow}
\usepackage{color}
\usepackage{latexsym}

\usepackage{url}
\usepackage{xcolor}
\definecolor{newcolor}{rgb}{.8,.349,.1}

\usepackage{hyperref}
\usepackage[final]{pdfpages}

\journal{Computers \& Graphics}

\begin{document}

\verso{Preprint Submitted for review}

\begin{frontmatter}

\title{Alternately denoising and reconstructing unoriented point sets}%

\author[1,3] {Dong Xiao}
\ead{xiaod18@mails.tsinghua.edu.cn}

\author[2,4]{Zuoqiang Shi}
\ead{zqshi@tsinghua.edu.cn}

\author[1,3] {Bin Wang\corref{cor}}
\ead{wangbins@tsinghua.edu.cn}
\cortext[cor]{Corresponding author at: School of Software, Tsinghua University, Beijing, China}

\address[1]{School of Software, Tsinghua University, Beijing, China}
\address[2]{Yau Mathematical Sciences Center, Tsinghua University, Beijing, China}
\address[3]{Beijing National Research Center for Information Science and Technology, Beijing, China}
\address[4]{Yanqi Lake Beijing Institute of Mathematical Sciences and Applications, Beijing, China}


\begin{abstract}

We propose a new strategy to bridge point cloud denoising and surface reconstruction by alternately updating the denoised point clouds and the reconstructed surfaces. In Poisson surface reconstruction, the implicit function is generated by a set of smooth basis functions centered at the octnodes. When the octree depth is properly selected, the reconstructed surface is a good smooth approximation of the noisy point set. Our method projects the noisy points onto the surface and alternately reconstructs and projects the point set. We use the iterative Poisson surface reconstruction (iPSR) to support unoriented surface reconstruction. Our method iteratively performs iPSR and acts as an outer loop of iPSR. Considering that the octree depth significantly affects the reconstruction results, we propose an adaptive depth selection strategy to ensure an appropriate depth choice. To manage the oversmoothing phenomenon near the sharp features, we propose a $\lambda$-projection method, which means to project the noisy points onto the surface with an individual control coefficient $\lambda_{i}$ for each point. The coefficients are determined through a Voronoi-based feature detection method. Experimental results show that our method achieves high performance in point cloud denoising and unoriented surface reconstruction within different noise scales, and exhibits well-rounded performance in various types of inputs. The source code is available at~\url{https://github.com/Submanifold/AlterUpdate}.

\end{abstract}


\end{frontmatter}

\section{Introduction}

Point clouds are widely applied in a wide range of geometric applications. However, the real scanned point clouds obtained by the sensing technologies typically contain a certain amount of noise and outliers, which significantly reduces their shape representation capacities. When the input point clouds are unoriented (i.e., no consistently oriented normals are given), applying them for surface reconstruction becomes a challenging task. Point cloud denoising is a traditional solution for handling low-quality inputs, which has been extensively studied for more than two decades. The denoising techniques can be classified into several categories. A typical idea is to project the noisy point set onto the estimated local surfaces. However, existing projection-based methods either require consistently oriented normals for robust shape fitting (e.g., MLS-projection~\cite{2003PointSetSurface,2009KernelRegre}) or only considers the local shape properties within a specific neighbor scale (e.g., jet smoothing~\cite{2005Jets, 2008Jetcpp}).

In recent years, researchers have made remarkable progress in unoriented surface reconstruction such as VIPSS~\cite{2019VIPSS}, PGR~\cite{2022pgr} and iPSR~\cite{2022ipsr}. Among them, iPSR iteratively performs screened Poisson surface reconstruction~\cite{2013SPSR} and updates the point normals from the surface generated by the previous iteration. The algorithm terminates when approximating the iterative fixed point or reaching the maximum number of iterations. iPSR achieves high performance for clean inputs. However, the sample positions are fixed during the iterative process, reducing its performance in point clouds with large noise. Therefore, we wonder if the point positions can also be updated to achieve a higher reconstruction quality.

In this work, we bridge point cloud denoising and surface reconstruction by alternately updating the denoised point clouds and the reconstructed surfaces. Our method acts as an outer loop of iPSR. Specifically, we project the noisy point set onto the surface generated by iPSR in each iteration. This idea is based on the following observations: in Poisson surface reconstruction, the reconstructed surface is generated by the isosurfacing of an implicit function spanned by a series of smooth basis centered at the octnodes. When a proper octree depth is selected, the reconstructed surface can be regarded as a good smooth approximation of the noisy point cloud. Furthermore, we notice that the depth selection will significantly affect the convergence of iPSR. Accordingly, we propose an adaptive depth selection method based on the convergence situation of iPSR, which is judged by the normal variations in the algorithm. To address the oversmoothing phenomenon in the sharp edges, we propose a $\lambda$-projection method by estimating the sharpness ratio and assigning an individual control coefficient for each point. The coefficients are determined using a Voronoi-based feature estimation algorithm~\cite{2011VoronoiFeature}. 

From the point cloud denoising viewpoint, our method belongs to the projection-based category and iteratively projects the noisy point set onto the surface generated by iPSR with an adaptive depth selection strategy and a $\lambda$-projection method. From the unoriented surface reconstruction viewpoint, our method iteratively updates the point positions in the algorithm (instead of updating the normals only) and achieves considerable improvements for noisy inputs. 

We qualitatively and quantitatively examine the efficacy of our method in point-wise noise, structured noise, outliers and real scanned noise. The experimental results indicate that our method achieves high performance in point cloud denoising and unoriented surface reconstruction tasks, and shows well-rounded performance within different shapes and noise scales. We also verify that our approach can not be substituted by simply adjusting some parameters in iPSR or other denoising and reconstructing approaches.

\section{Related works} \label{related_work}

In this section, we provide a brief review of point cloud denoising and unoriented surface reconstruction techniques and focus on the works that are most relevant to our method.

\subsection{Projection-based point cloud denoising} \label{projection-based}

Point cloud denoising has been extensively studied for more than two decades due to its broad applications. Han et al.~\cite{2017ReviewFilter} provide a systematic review involving a wide variety of traditional approaches. A recent survey paper~\cite{2022PointCloudReview} introduces more up-to-date techniques including many interesting learning-based methods. Projection-based denoising is the most relevant category of our method with the main idea to project the noisy point cloud onto the local surfaces approximated by adjacent points. A commonly used local fitting technique is the Moving Least Squares (MLS) surfaces. The seminar work~\cite{2003PointSetSurface} defines a $C^{2}$ smooth surface through the projection procedure. However, the sharp features may be oversmoothened due to the smoothness property. Fleishman et al.~\cite{2005MovingSharp} approximate the point set by the feature-preserving piecewise smooth surfaces. Each piece fits the point shape on one side of the sharp edge. APSS~\cite{2007APSS} utilizes algebraic spheres to define the MLS surfaces, which significantly improves the quality near the high curvature regions. RIMLS~\cite{2009KernelRegre} applies the robust local kernel regression to achieve feature preserving surface approximation. Majority of the above-mentioned techniques require consistently oriented normals as inputs. APSS supports both unoriented and oriented point fitting. However, normals are necessary for achieving the high-fidelity quality. Except for MLS projection, Cazals et al.~\cite{2005Jets, 2008Jetcpp} estimate local shape properties through the Taylor expansion of the local height field. Then, a denoising technique can be realized by projecting the noisy points onto the approximated local surface. The aforementioned method is named as ``jet smoothing", which does not require normals as inputs. However, the local fitting is only performed within a user specific neighbor scale.

Except for local surface fitting, some other approaches carry out point projection through an optimization manner. Lipman et al.~\cite{2007LOP} propose a parameterization-free locally optimal projection (LOP) operator, which generates a set of uniform distributed points to approximate the original shape. This technique is further improved by Huang et al.~\cite{2009WLOP} through a robust repulsion term and the adaptive density weights. Liu et al.~\cite{2012iterconsolidation} suggest performing WLOP in an iterative manner. Preiner et al.~\cite{2014CLOP} propose a continuous WLOP formulation based on the Gaussian mixture technique. This work significantly reduces the time resources of the WLOP. However, the above-mentioned methods may lack the feature preserving mechanisms and result in oversmoothing near the sharp edges. To address this issue, Huang et al.~\cite{2013EdgeAware} propose to detect the sharp regions and carry out a point resampling operation near the sharp edges. Lu et al.~\cite{2018GMMFilter} propose a GMM-inspired anisotropic projection strategy, which preserves the sharp features by considering filtered normals in the objective function.

\subsection{Other point cloud denoising techniques}\label{other_denoising}

Projection is not the only means to carry out point cloud denoising. Digne et al.~\cite{2017Bilateral} and Zhang et al.~\cite{2019BilateralPCA} introduce the bilateral filters for 3D point set filtering. Their methods are mainly inspired by image denoising techniques. Avron et al.~\cite{2010l1sparse} and Sun et al.~\cite{2015l0sparse} propose the L1 and L0 optimization formulation for point set filtering based on the observation that many surfaces are piecewise smooth. Accordingly, the sparsity of the first order information can be applied to clean the point set. Agathos et al.~\cite{2022Elliptic} extend the Taubin smoothing techniques from the mesh denoising to point clouds, and implement a fast GPU version to support large scale data. Salman et al.~\cite{2010meshgeneration} detect sharp features by analyzing the covariance matrices of Voronoi cells. Digne et al.~\cite{2014simplification} utilize an error metric based on optimal transport to achieve the feature-capturing reconstruction and simplification. Wang et al.~\cite{2013CGFWangetal} present a feature-preserving unoriented reconstruction pipeline and separate the entire task into several processes, including noise scale estimation, tangent plane detection, outlier removal, feature detection and noise smoothing. Liu et al.~\cite{2020featurepreserve} propose a feature detection technique by the bi-tensor voting scheme. Chen et al.~\cite{2023Unifilter} propose to incorporate a repulsion term with the data term and take into account both point distribution and feature preservation. RFEPS~\cite{2022RFEPS} introduces a novel feature-line detection and resample approach for CAD models. 

The application of deep neural networks to geometric processing has achieved significant success in recent years due to the rapid advancement of 3D deep learning. Some interesting works learn the local shape properties through point-based 3D networks and obtain considerable results in point cloud denoising and unoriented surface reconstruction~\cite{2018EC-Net, 2020Pointcleannet, 2020P2S, 2020ConvPoint, 2021Pointfilter, 2022MODNet, 2023PCDNF, 2023IterativePFN}. However, the above-mentioned techniques typically require abundant training data to learn the shape priors. 

\begin{figure*}[htb]
  \centering
  \includegraphics[width=1.0\linewidth]{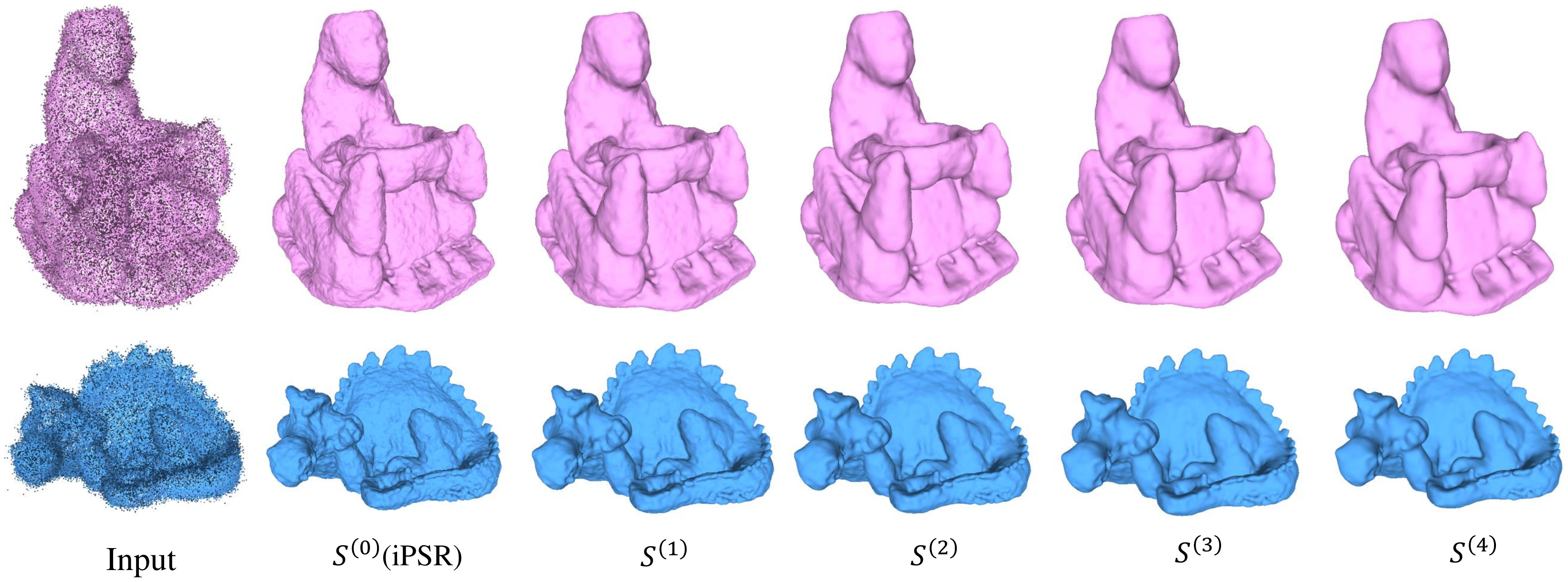}
  \caption{\label{fig:Figure1}
           Given the input point cloud $P^{(0)}$, our method iteratively performs reconstruction and projection in every step. We show the reconstructed surfaces from $S^{(0)}$(iPSR) to $S^{(4)}$ of our method. The surface quality is improved through the iteration.}
\end{figure*}

\subsection{Surface reconstruction from unoriented point sets} \label{point_recon}

Despite a long research history with a considerable number of excellent studies~\cite{2007VoronoiRecon, 2010Signing, 2013NoiseAdaptive, 2019VIPSS}, surface reconstruction from unoriented point clouds remains a challenging problem. The main reason is that obtaining a globally consistent normal orientation is always a non-trivial task. Some remarkable studies have emerged in recent years and achieve high-fidelity reconstruction results for clean inputs~\cite{2022pgr, 2022ipsr}. However, the point positions are fixed during the algorithm, limiting their effects on point clouds with a certain amount of noise. This is why we propose to fill this gap by updating the point positions iteratively. In contrast with these methods, DPSR~\cite{2021DPSR} proposes a differentiable Poisson solver and updates the source point cloud through the backpropagation of the Chamfer loss. However, the computational process of DPSR relies on the GPU, which results in the high resource demand and the restricted grid resolution. 

\section{Method} \label{Method}
\subsection{Overview} \label{Overview}

Given an input 3D point cloud $P^{(0)}=\{p_1, p_2,...,p_n\}$ with noise, our method iteratively updates the reconstructed surface and the denoised point cloud. The two main operators in our algorithm are the reconstruction operator and the projection operator. We apply the iterative Poisson surface reconstruction (iPSR)~\cite{2022ipsr} to carry out unoriented surface reconstruction, which takes a raw point cloud $P$ and a user-specific octree depth $d$ as inputs, and generates a smooth surface $S$ directly from the point positions. This operation can be denoted as follows:
\begin{equation}
\label{equation1}
S = f_{ipsr}(P, d).
\end{equation}
Then, the noisy points are projected onto the surface. Instead of directly projecting the points onto $S$, we propose a $\lambda$-projection operator $f_{proj}$, which takes a point cloud $P$ and a surface $S$ as inputs and projects each point $p_i\in P$ on $S$ with a control coefficient $\lambda_{i}$. If $q_i$ is the closest point to $p_i$ on $S$, then the projected point $p_{i}^{\prime}$ satisfies:
\begin{equation}
\label{equation2}
p_{i}^{\prime} = (1 - \lambda_{i})p_{i} + \lambda_{i} q_{i}.
\end{equation}
This projection operation is denoted as follows:
\begin{equation}
\label{equation3}
P^{\prime} = f_{proj}(P, S).
\end{equation}
Generally, the alternative updating process of our method can be expressed as follows:
\begin{equation}
\label{equation4}
S^{(k)} = f_{ipsr}(P^{(k)}, d^{(k)}), \\
P^{(k+1)} = f_{proj}(P^{(k)}, S^{(k)}).
\end{equation}

Our method acts as an outer loop of iPSR. Fig.~\ref{fig:Figure1} shows the input point cloud $P^{(0)}$ and the reconstructed surfaces generated by our method from $S^{(0)}$(iPSR) to $S^{(4)}$. It can be seen that the mesh quality is improved and the noise is cleaned through the iteration.

\begin{figure}[htb]
  \centering
  \includegraphics[width=1.0\linewidth]{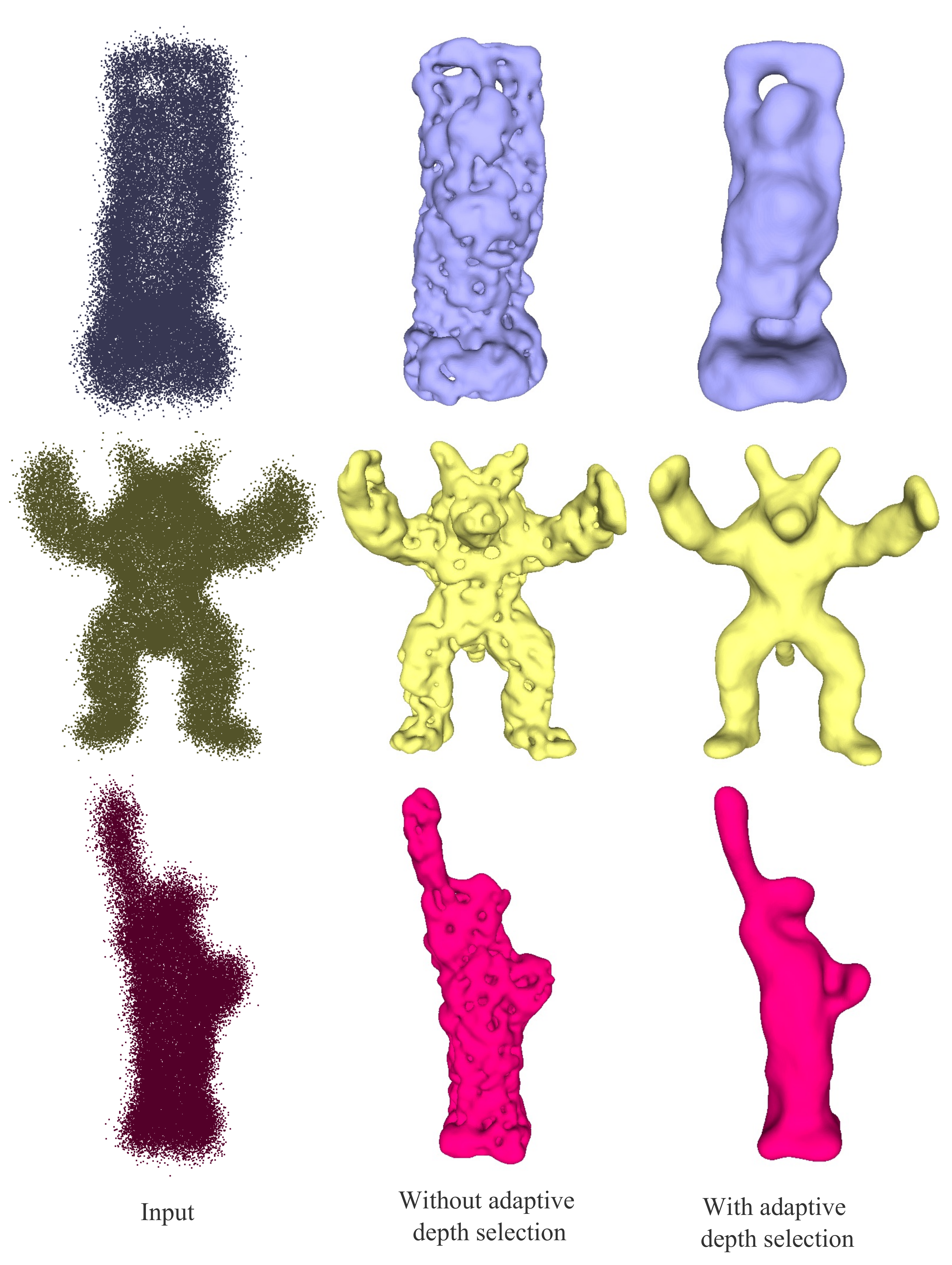}
  \caption{\label{fig:Figure2}
           The effectiveness of the adaptive depth selection strategy in large noise examples. If the depth is simply set to 8, as shown in the middle column, then the resulting meshes contain ``lattice'' structures. By contrast, utilizing our adaptive depth selection method brings about significant improvements in these examples.}
\end{figure}

In section~\ref{AdaptiveDepth}, we will introduce the adaptive octree depth selection strategy to determine the depth $d^{(k)}$ in Equation~\ref{equation4}. In section~\ref{lambdaprojection}, we will introduce how the projection coefficient $\lambda_i$ is selected for each point to handle the oversmoothing phenomenon near the sharp edges.

\subsection{Adaptive depth selection strategy} \label{AdaptiveDepth}
When carrying out accurate and high-fidelity reconstruction by the traditional Poisson surface reconstruction approach~\cite{2006PSR, 2013SPSR}, the octree depth is always set to a large value for supporting detailed isosurfacing. In the original conception, a larger octree depth would lead to a better mesh quality, until reaching the maximum depth required for the geometric complexity. However, this situation is not the case when doing unoriented reconstruction for noisy point clouds using iPSR. Firstly, we will provide a brief review of the normal iteration process of iPSR as preliminaries.

\textbf{iPSR.} Given an unoriented point cloud $P$ and a user specific octree depth $d$, iPSR firstly constructs an octree from $P$ and generates a new sample set $\mathcal{S}=\{s_1, s_2,...,s_m\}$ based on the octree nodes. Each sample $s_{i}$ is assigned with an initialized normal $n_{i}$. The normals can be randomly initialized in the first iteration. We denote the set of normals as $\mathcal{N}=\{n_1, n_2,...,n_m\}$. iPSR iteratively updates the sample normals $\mathcal{N}$ from the surface generated by the last iteration. Specifically, an intermediate surface is generated by inputting $\mathcal{S}$ and $\mathcal{N}$ to the screened Poisson surface reconstruction~\cite{2013SPSR}. Then, the normal $n_{i}$ of the sample $s_i$ is updated by the average normal of the adjacent face list near $s_i$ in the intermediate surface. The algorithm terminates when the iterative fixed point is approximated or the maximum iteration is reached. Then, a consistent and high quality triangular mesh can be obtained from the last iteration. 

In the original work of iPSR, the octree depth is manually set to a fixed integer 10 in most experiments. However, we notice that choosing an appropriate depth for large noise inputs can bring about considerable advantages for this algorithm. When the octree depth is considerably large, the convergence of iPSR may encounter some challenges for large noise inputs and produce ``lattice" surfaces, such as the middle column of Fig.~\ref{fig:Figure2}. Reducing the octree depth can make the convergence easier and produce smoother surfaces. However, an extremely small depth will result in the loss of the shape details, as shown in the teapot mouth of Fig.~\ref{fig:Figure4b} when doing reconstruction of depth 6. Therefore, an appropriate depth is critical for achieving a decent performance.

In this work, we propose an adaptive depth selection strategy based on the convergence situation (i.e., good or bad) of iPSR. Recall that iPSR updates the sample normals in an iterative manner. In their algorithm, the average of the top 0.1\% normal variations $v$ is calculated for each iteration. This value can be applied to judge if the algorithm has well converged. The iteration process of iPSR stops when $v<0.175$, or the maximum iteration $30$ is reached. Accordingly, we also apply the normal variations to estimate the convergence situation of iPSR in the first reconstruction of our method. Specifically, we set the candidate depths to be $[d_{min}, d_{max}]$ and first perform iPSR by depth $d = d_{max}$. If iPSR converges and stops before reaching the maximum normal iteration $30$, or the average normal variations of the last five iterations are less than $0.7$, then iPSR converges well at this depth and $d^{(0)} = d$ is set. Otherwise, $d = d - 1$ and iPSR is performed again, until converges well or reaches the minimum depth $d = d_{min}$. This approach ensures that a proper depth is chosen at the first reconstruction.

The depth $d^{(0)}$ of the first reconstruction accounts for a critical role in our algorithm, and has been determined through the  above-mentioned strategy. In the subsequent reconstructions, we can set an incremental depth list for each $d^{(0)}\in[d_{min}, d_{max}]$. For instance, the depth list is set to $[6,6,7,7,8]$ for $d^{(0)}=6$. The depth increases in the following stages because the noise magnitude becomes smaller through the alternative denoising and reconstructing process. Additional details will be specified in the experimental section. Once $d_{min}$ and $d_{max}$ are determined, and the depth list for each intermediate depth $d\in[d_{min}, d_{max}]$ is set, we can run the dataset in batches without manually adjusting the depth for each shape or reconstruction in Equation~\ref{equation4}. Therefore, the depth selection strategy of our method is adaptive.

\subsection{$\lambda$-projection method} \label{lambdaprojection}
In each iteration of our method with the input point cloud $P^{(k)}$ and the reconstructed surface $S^{(k)} = f_{ipsr}(P^{(k)}, d^{(k)})$, we project each point $p_{i}^{(k)} \in P^{(k)}$ onto the reconstructed surface $S^{(k)}$. However, if we simply project all the points onto the surface, then the sharp edges will be smoothed into rounded corners after several iterations. Accordingly, a $\lambda$-projection method is proposed to combat the issue of oversmoothing near the sharp edges. The main idea is to assign a projection coefficient $\lambda_{i}^{(k)}$ for each point $p_{i}^{(k)}$. If $q_{i}^{(k)}$ is the nearest point of $p_{i}^{(k)}$ in the surface $S^{(k)}$, then the projected points $p_{i}^{(k+1)}$ satisfy:
\begin{equation}
\label{equation7}
p_{i}^{(k+1)} = (1 - \lambda_{i}^{(k)})p_{i}^{(k)} + \lambda_{i}^{(k)} q_{i}^{(k)}.
\end{equation}
The coefficient $\lambda_{i}^{(k)}$ is different for each point and mainly depended on its sharpness degree. In this work, a Voronoi-based feature estimation method~\cite{2011VoronoiFeature} is used to examine the sharpness degree of all the points. Given the point cloud $P^{(k)}=\{p_{i}^{(k)}\}_{i=1}^{n}$, ~\cite{2011VoronoiFeature} firstly computes the convolved Voronoi covariance measures from the neighborhood of each $p_{i}^{(k)}$. Then, the sharpness ratio $r_{i}^{(k)}$ can be computed from the eigenvalues of the covariance measure. The point whose sharpness ratio is larger than a certain threshold is considered as a feature point in the shape.

Once the sharpness ratios of the point set are calculated, the projection coefficient $\lambda_{i}^{(k)}$ can be determined from the sharpness ratio $r_{i}^{(k)}$. We set $\lambda_{i}^{(k)}$ to be a continuous function of $r_{i}^{(k)}$:
\begin{equation}
\label{equation8}
\lambda_{i}^{(k)} = 0.1 + 0.9 \times e^{-g(r_{i}^{(k)})},
\end{equation}
where
\begin{equation}
\label{equation8b}
g(r_{i}^{(k)}) = \frac{{(max \{r_{i}^{(k)} - c, 0\})}^{2}}{\sigma^{2}}.
\end{equation}
Here, $c$ and $\sigma$ are both user specific parameters, which represent the threshold and the standard deviation, respectively. Equation~\ref{equation8} and~\ref{equation8b} demonstrate that $\lambda_{i}^{(k)} \in (0.1, 1.0]$. If $r_{i}^{(k)} < c$, then $p_{i}^{(k)}$ is not regarded as a feature point, $\max \{r_{i}^{(k)} - c, 0\} = 0$, and $\lambda_{i}^{(k)} = 1.0$, indicating that point $p_{i}^{(k)}$ is projected onto the surface at $q_{i}^{(k)}$ ($p_{i}^{(k+1)}=q_{i}^{(k)}$). If $r_{i}^{(k)} > c$, then $p_{i}^{(k)}$ is regarded as a feature point with a large degree of sharpness. As the sharpness ratio $r_{i}^{(k)}$ increases, $\lambda_{i}^{(k)}$ will decrease, but not less than the basic offset 0.1. $\sigma$ controls the decreasing speed of $\lambda_{i}^{(k)}$.

It should be noticed that the sharp edges are only considered when the noise scale is not that large. In section~\ref{AdaptiveDepth}, we propose an adaptive depth selection method. Large noise will result in a low $d^{(0)}$. Here, we choose a depth $d_{sharp}$ as threshold. When $d^{(k)} < d_{sharp}$, the sharp edges are not taken into account in this iteration because of the following reasons: 1) the input point cloud contains large noise; and 2) it is still in the early stage of the algorithm and the top priority of this iteration is to reduce the noise amplitude of the point cloud. When $d^{(k)} < d_{sharp}$, we set $\lambda_{i}^{(k)} = 0.5$ for all points, which not only filters the noise but also maintains the original shape. When $d^{(k)} \ge d_{sharp}$, Equation~\ref{equation8} and~\ref{equation8b} are utilized to calculate the projection coefficient $\lambda_{i}^{(k)}$ for each point.

\begin{figure}[htb]
  \centering
  \includegraphics[width=1.0\linewidth]{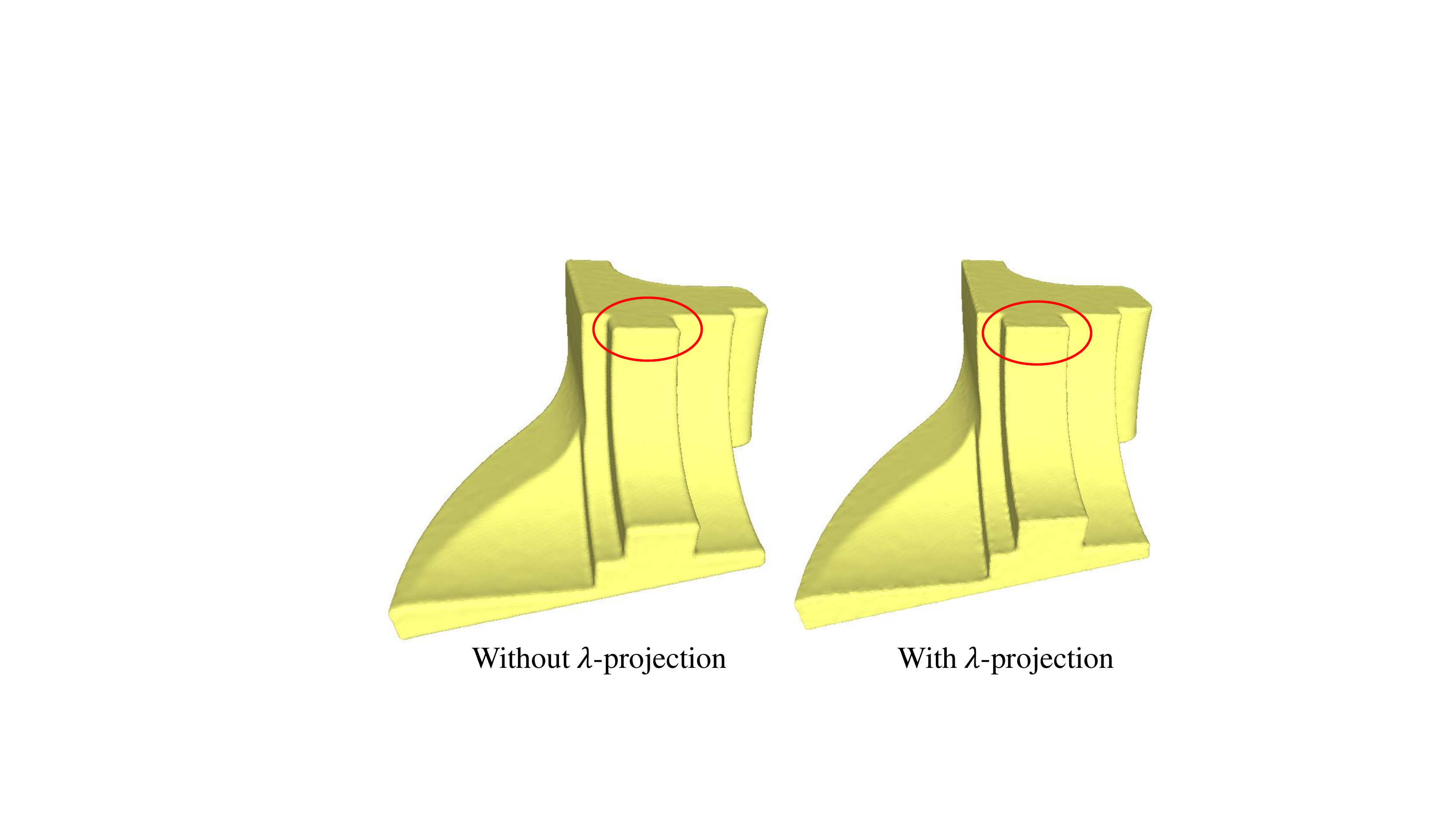}
  \caption{\label{fig:Figure3}
           Reconstructed mesh of a noisy \emph{fandisk} model with and without the $\lambda$-projection strategy. The results indicate that the proposed $\lambda$-projection method can ameliorate the oversmoothing phenomenon near the sharp edges.}
\end{figure}

In Fig.~\ref{fig:Figure3}, we show the reconstructed surface of a noisy \emph{fandisk} model with and without the $\lambda$-projection method. The results indicate that the proposed $\lambda$-projection method is helpful in ameliorating the oversmoothing phenomenon near the sharp edges. For further improvements to this issue, we can carry out a point resample process near the sharp edges similar to EAR~\cite{2013EdgeAware} and RFEPS~\cite{2022RFEPS}. 

The normal inputs of iPSR can be either random or manually initialized. Therefore, in the subsequent reconstructions of our method, we initialize the point normal of $p_{i}^{(k+1)}$ as the surface normal at $q_{i}^{(k)}$. In this way, the convergence speed of iPSR is significantly improved.

\begin{figure*}[htb]
  \centering
  \includegraphics[width=1.0\linewidth]{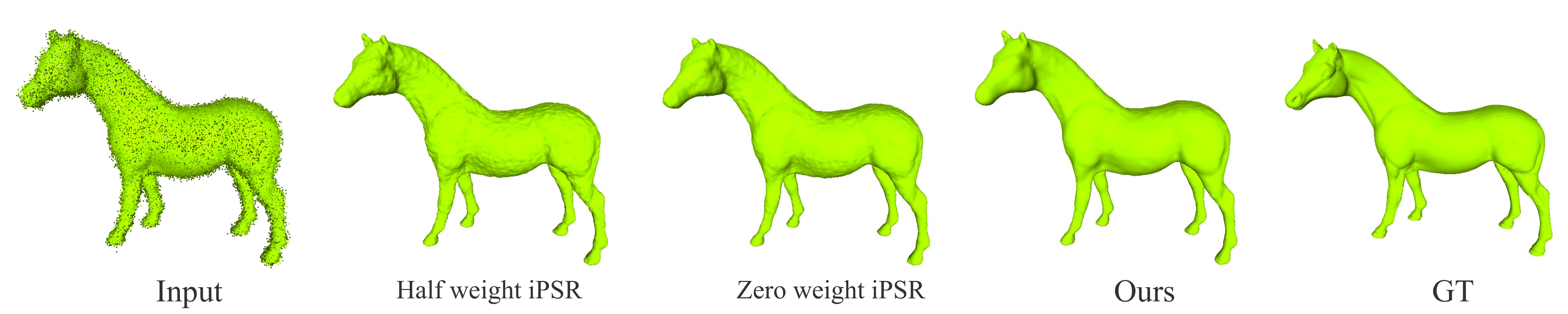}
  \caption{\label{fig:Figure4}
           Reconstruction of our method in point weight 1.0 and reconstructions of the iPSR in different smaller point weights. The results indicate that our method cannot be substituted by simply adjusting the point weight of iPSR.}
\end{figure*}

\begin{figure*}[htb]
  \centering
  \includegraphics[width=1.0\linewidth]{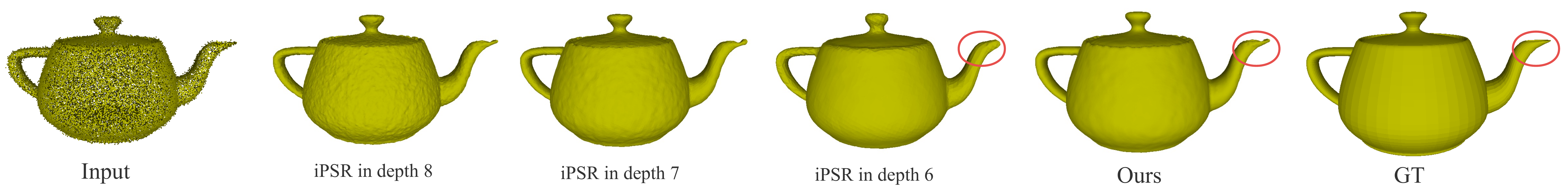}
  \caption{\label{fig:Figure4b}
           Comparisons of iPSR at different depths with our method. The results indicate that our method cannot be substituted by simply adjusting the depth of iPSR.}
\end{figure*}

\begin{figure*}[htb]
  \centering
  \includegraphics[width=1.0\linewidth]{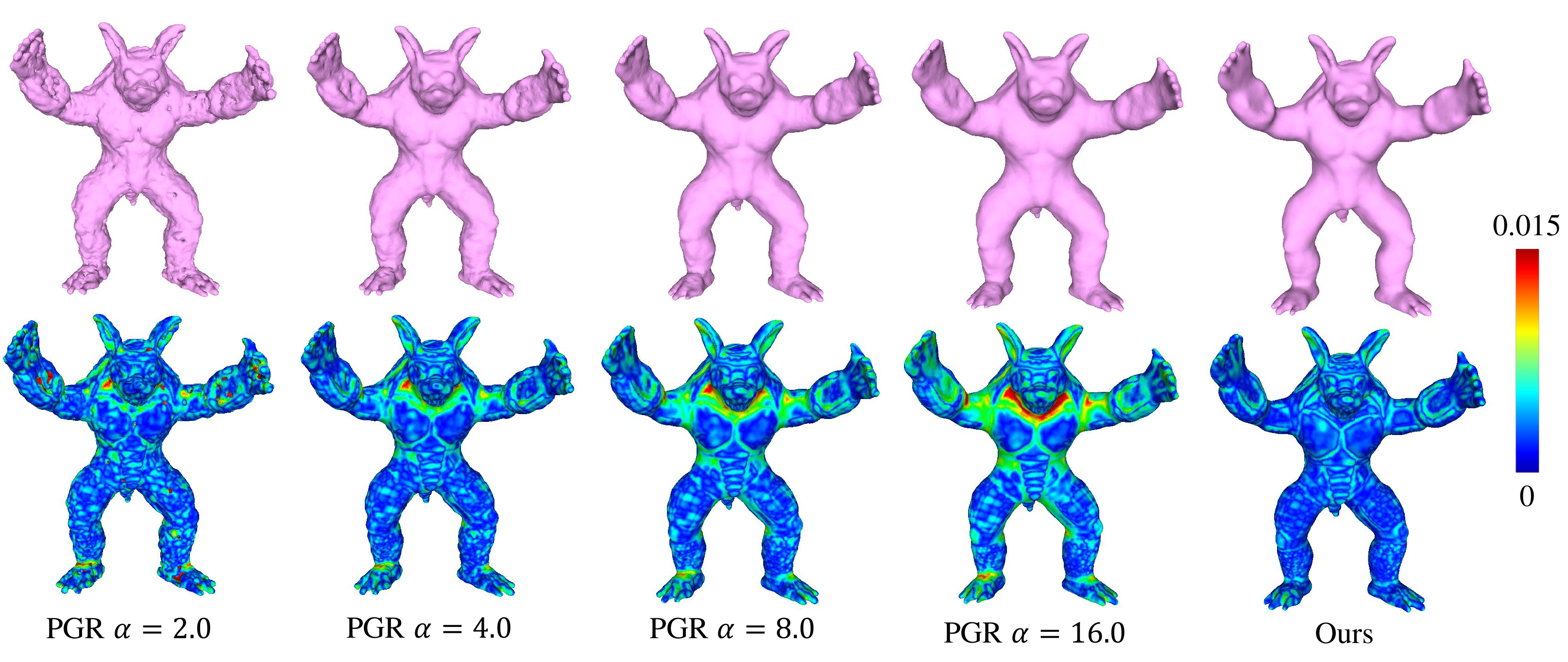}
  \caption{\label{fig:Figure12}
           Comparisons of PGR in different $\alpha$ values with our method. The results indicate that our method produces good visual effects and has a low distance error between the generated surface and the ground truth.}
\end{figure*}

\begin{figure*}[htb]
  \centering
  \includegraphics[width=1.0\linewidth]{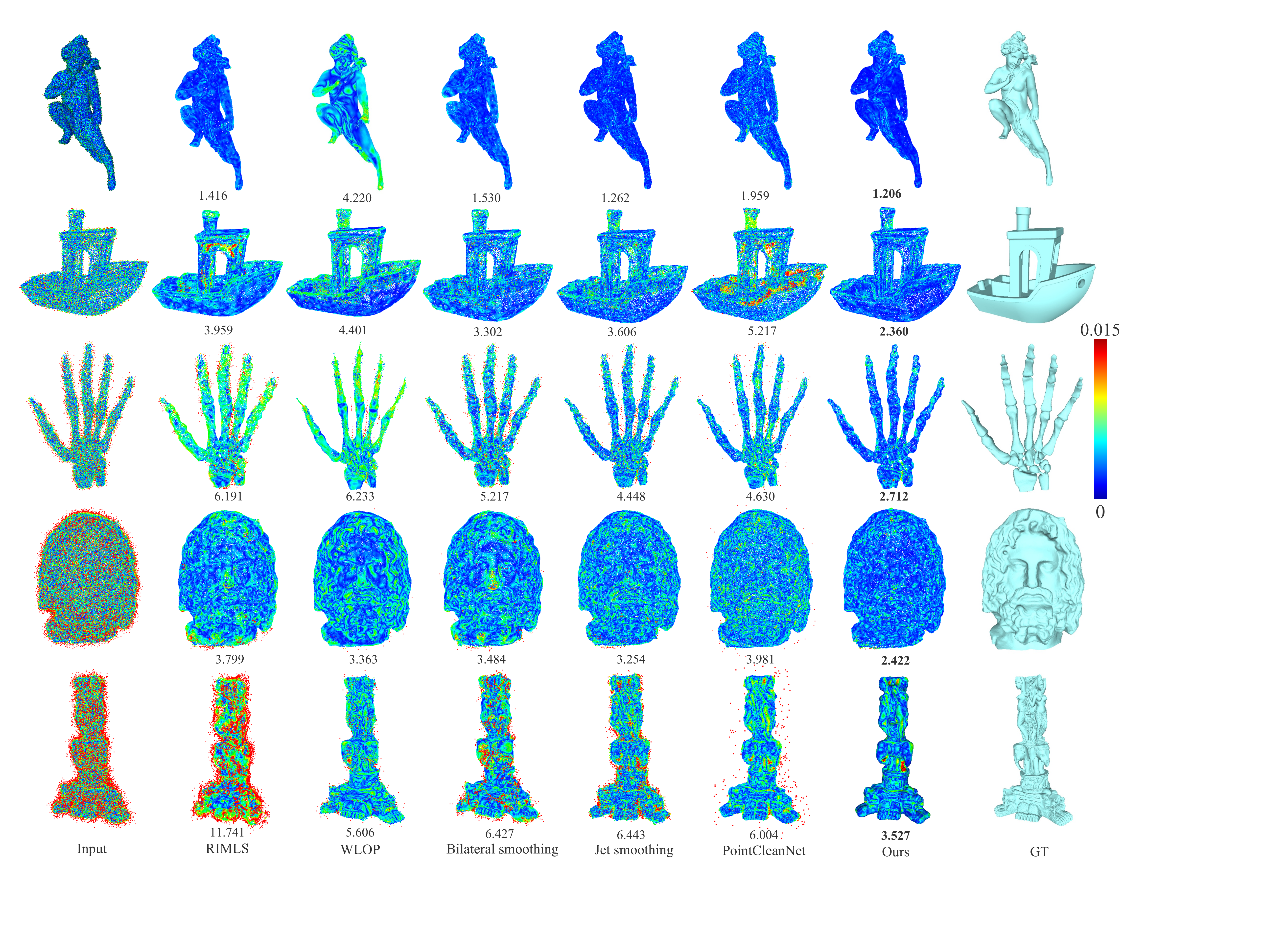}
  \caption{\label{fig:Figure5}
           Qualitative comparisons of the denoised point clouds. We annotate the RMSD value (multiplied by $10^{3}$) at the bottom of each example and colorize the point to surface distance from the denoised point cloud to the ground truth mesh.}
\end{figure*}

\begin{figure}[htb]
  \centering
  \includegraphics[width=1.0\linewidth]{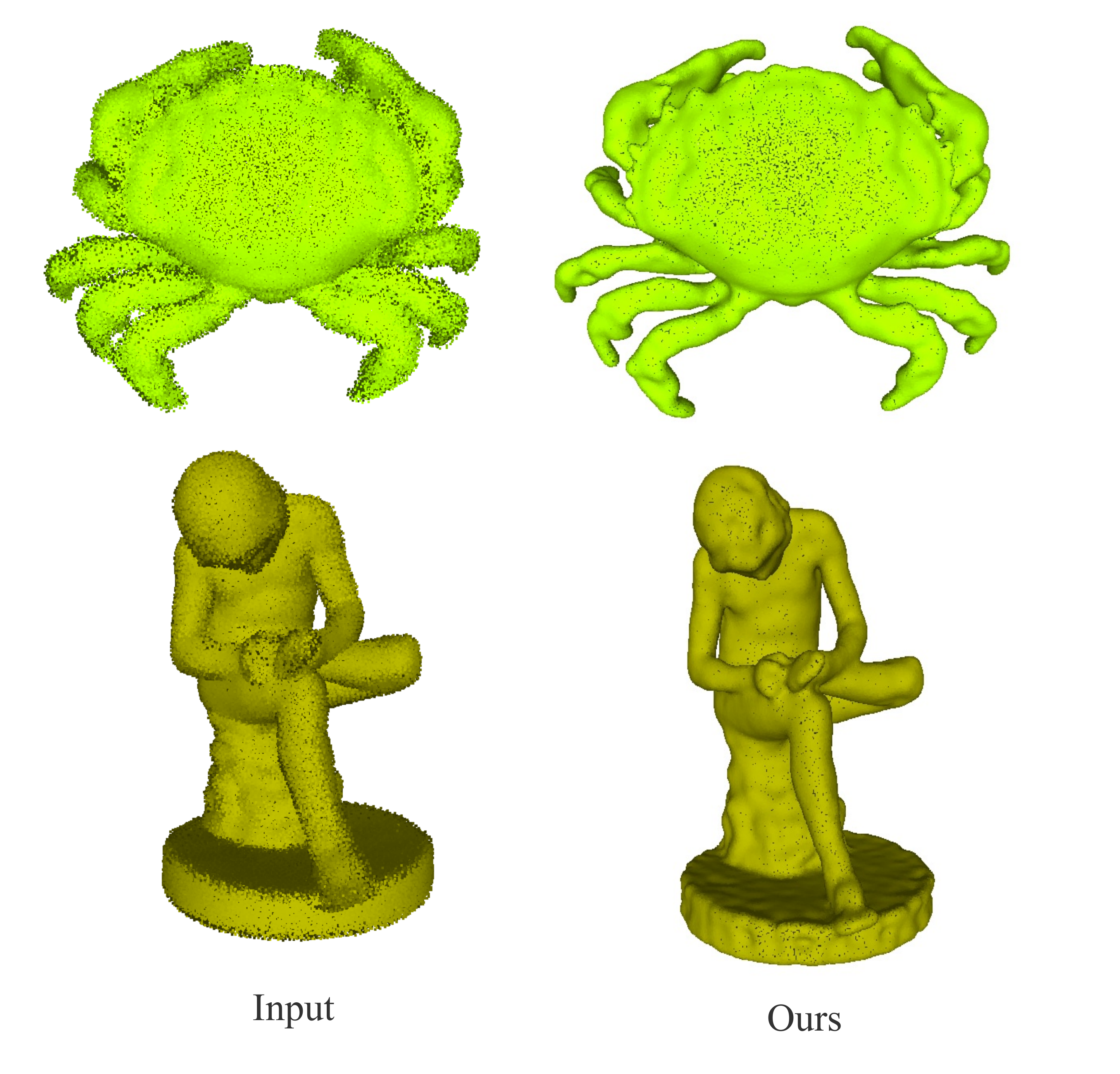}
  \caption{\label{fig:Figure10}
           Denoised results of our method in misalignment point clouds. The results indicate that our approach effectively addresses this situation.}
\end{figure}

\begin{figure}[htb]
  \centering
  \includegraphics[width=1.0\linewidth]{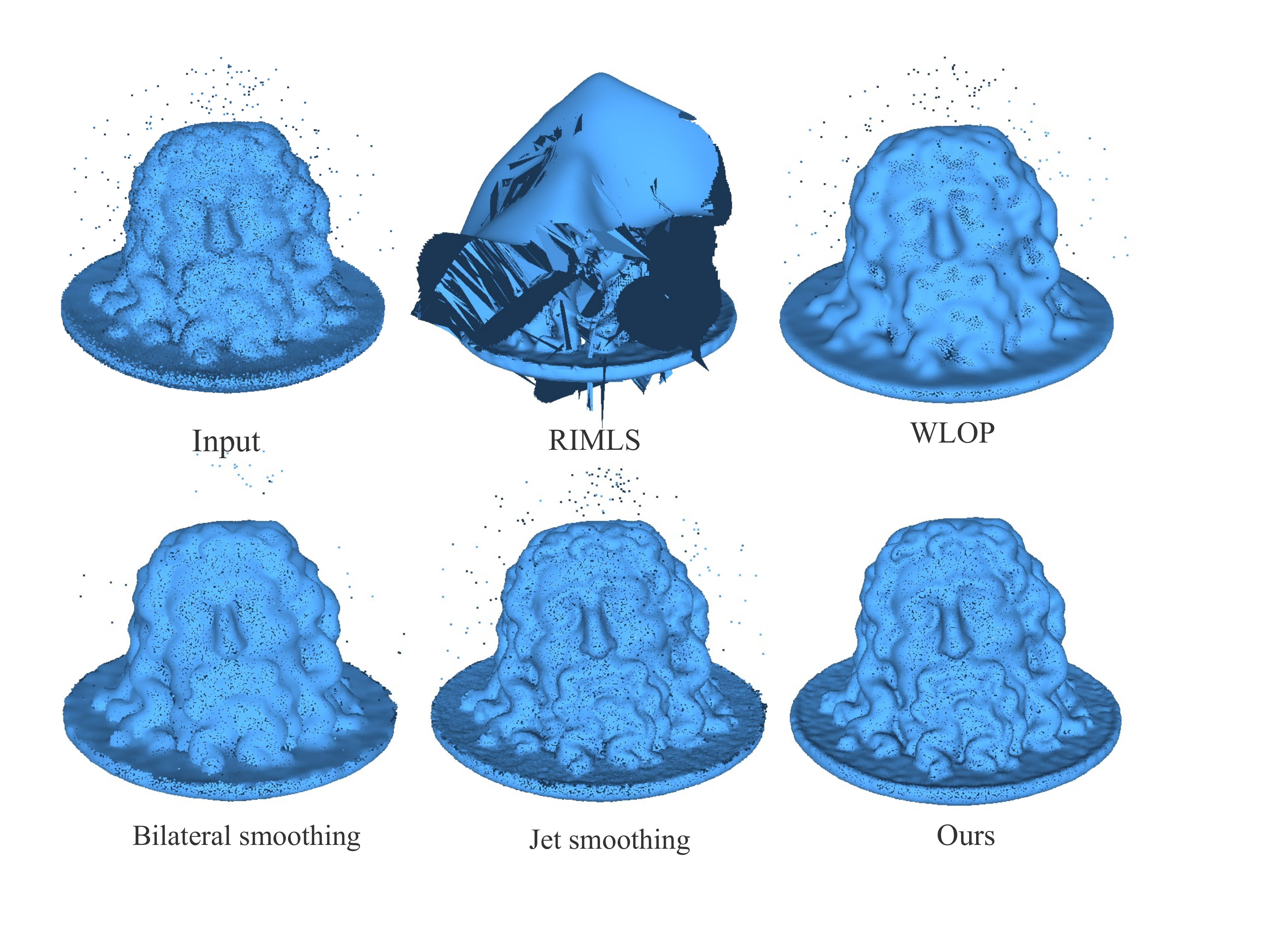}
  \caption{\label{fig:Figure6}
           The results amongst different methods of a point cloud contains both misalignment artifacts and outliers. Approximately $1$K points are randomly added in the unit cube as outliers. }
\end{figure}

\begin{figure}[htb]
  \centering
  \includegraphics[width=1.0\linewidth]{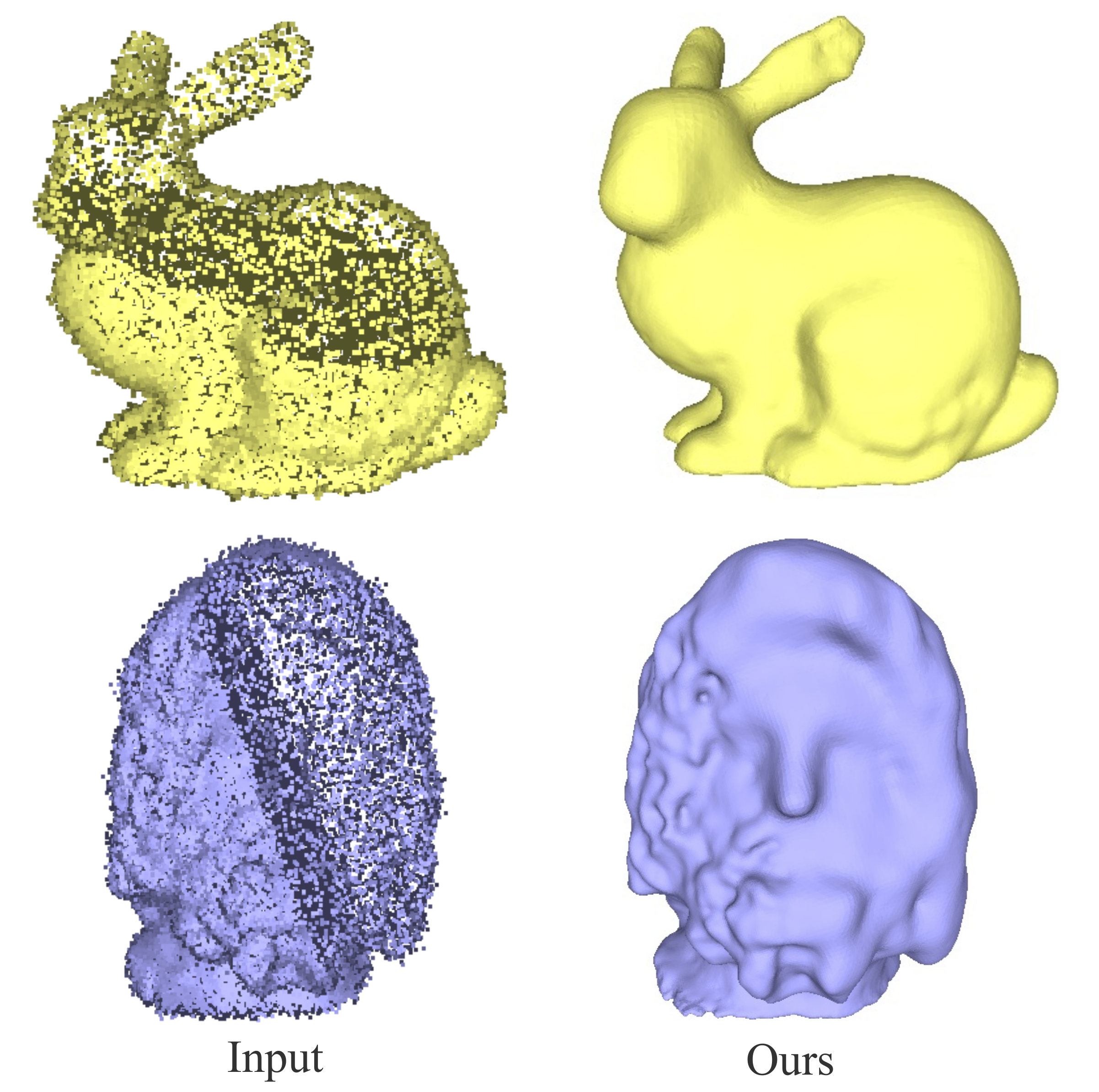}
  \caption{\label{fig:Figure14}
           Reconstructions of our method in point clouds with both noise and highly inconsistent sampling densities. Our approach effectively addresses this situation.}
\end{figure}

\begin{figure}[htb]
  \centering
  \includegraphics[width=1.0\linewidth]{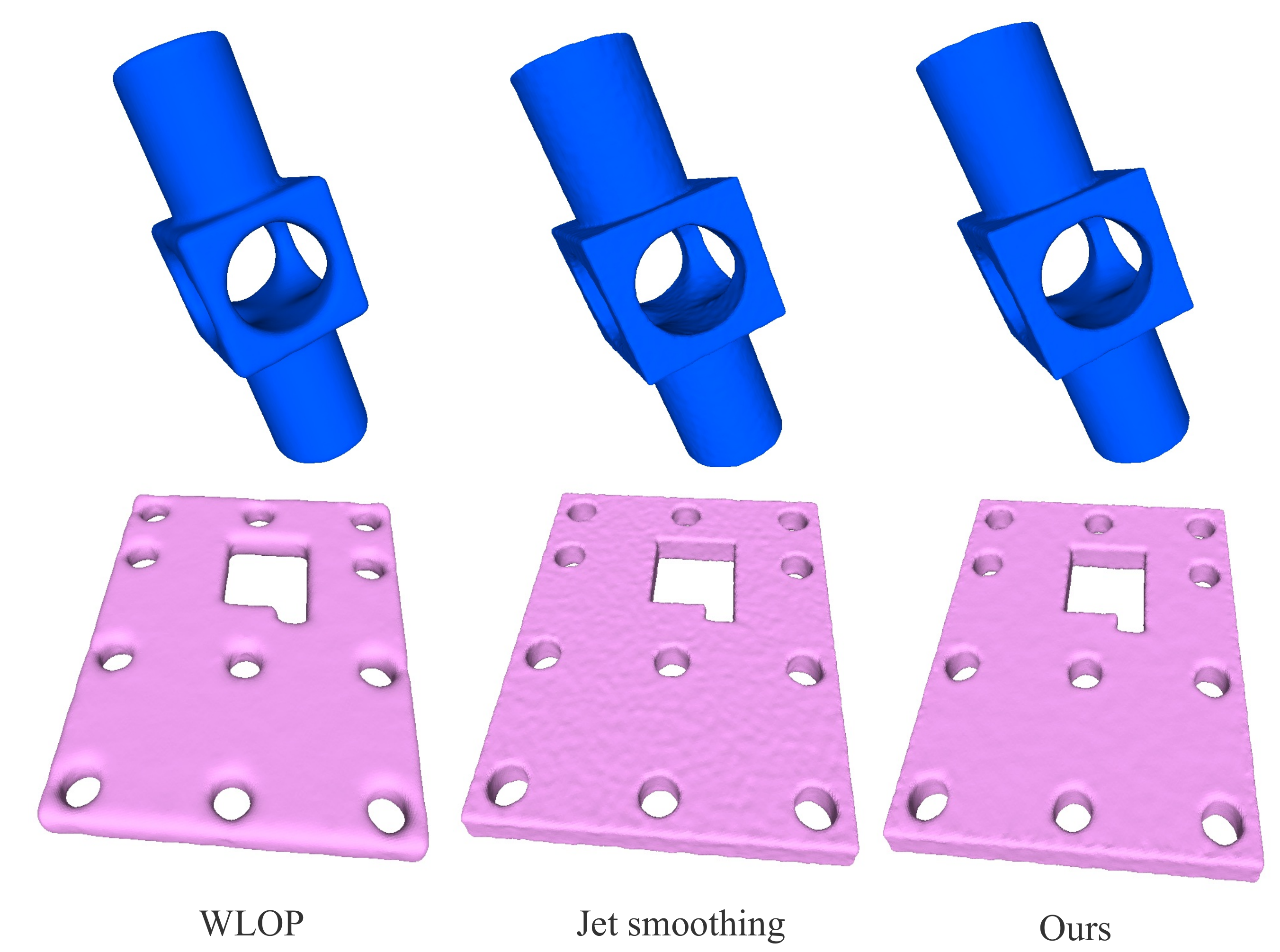}
  \caption{\label{fig:Figure7}
           Qualitative comparisons of our method with WLOP and jet smoothing in the CAD-like models. }
\end{figure}

\begin{figure}[htb]
  \centering
  \includegraphics[width=1.0\linewidth]{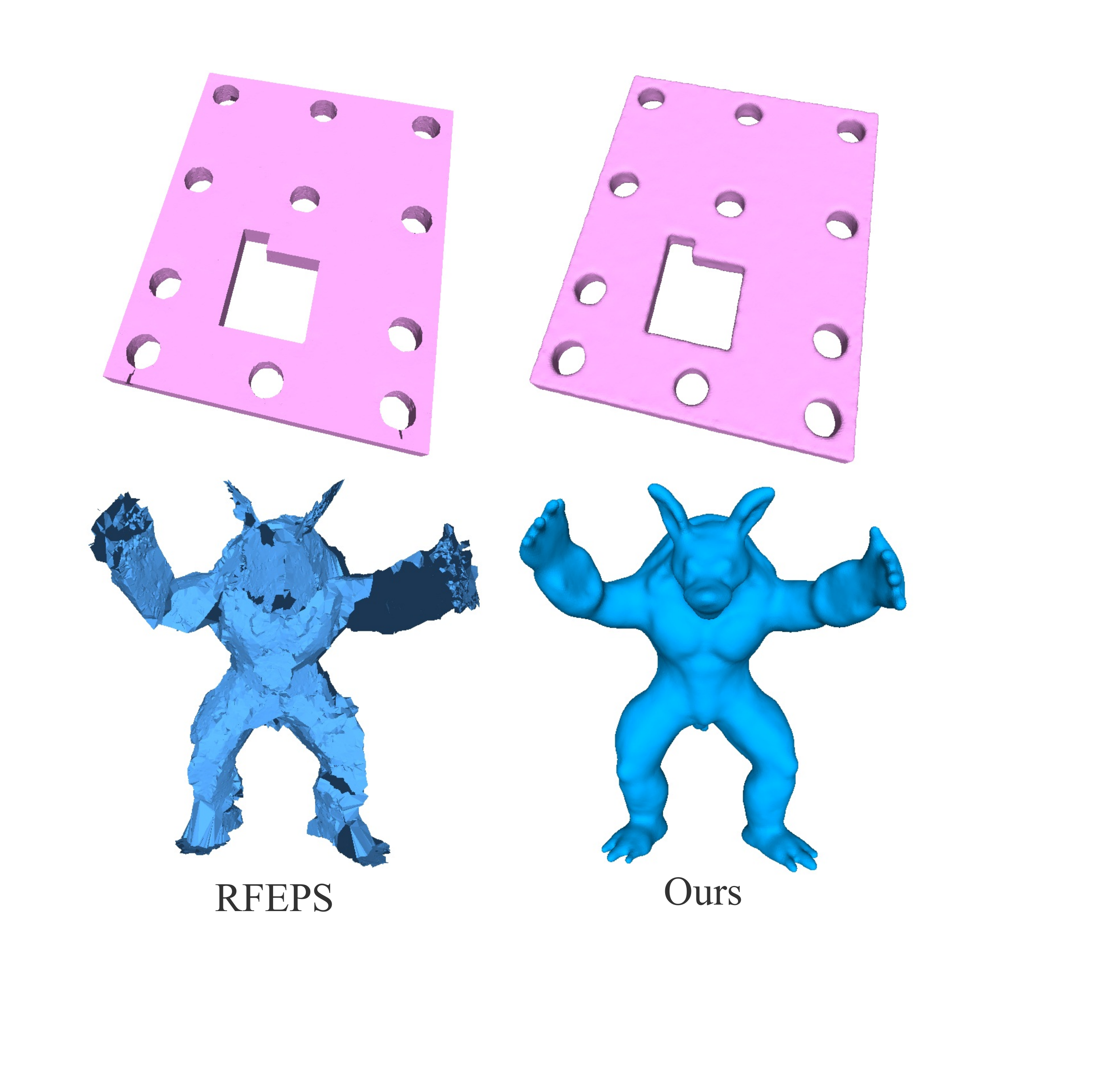}
  \caption{\label{fig:Figure13}
            Qualitative comparisons of our method with RFEPS. While RFEPS performs well in CAD models, its applicability to general shapes may be limited.}
\end{figure}

\begin{figure*}[htb]
  \centering
  \includegraphics[width=0.9\linewidth]{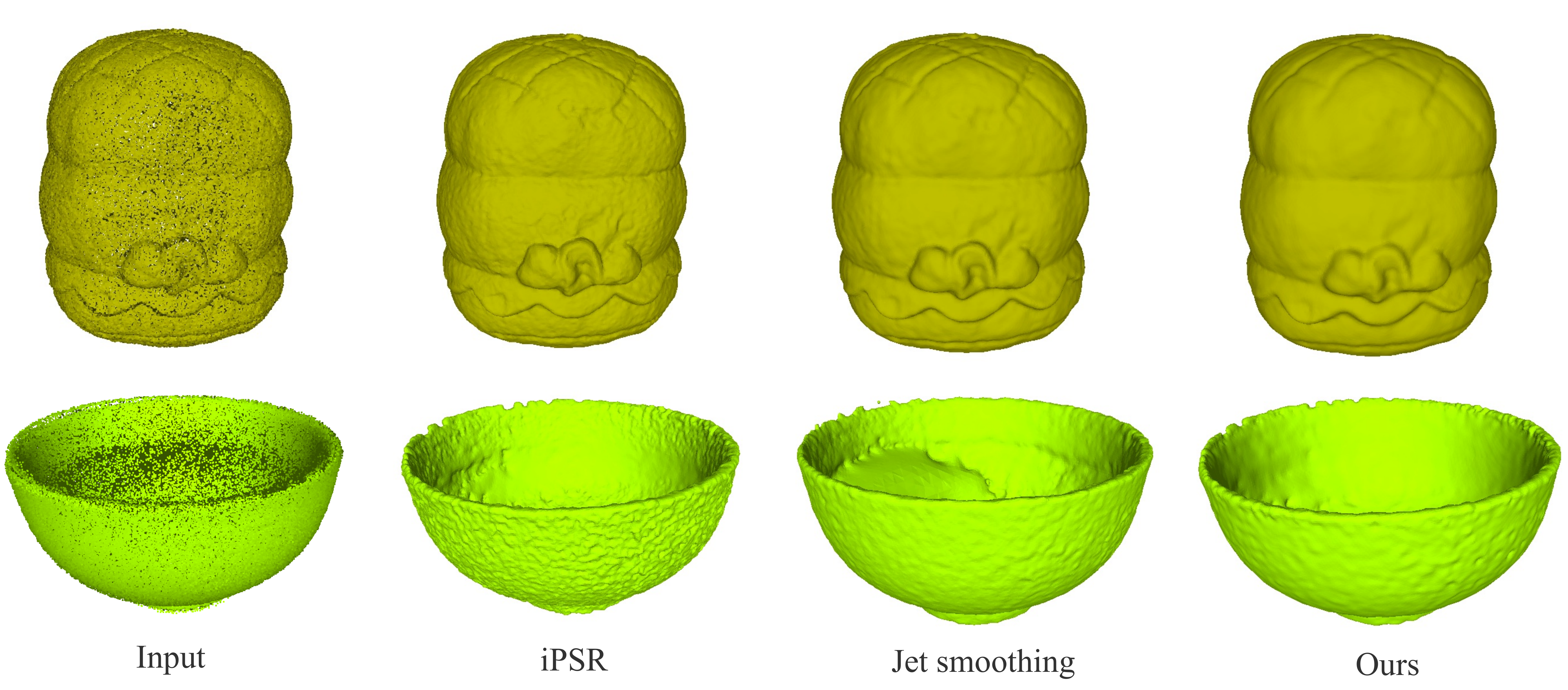}
  \caption{\label{fig:Figure9}
           Reconstruction results on the real scanned point clouds. We compare our method with iPSR and jet smoothing. Our method effectively manages the real scanned noise.}
\end{figure*}

\begin{figure}[htb]
  \centering
  \includegraphics[width=1.0\linewidth]{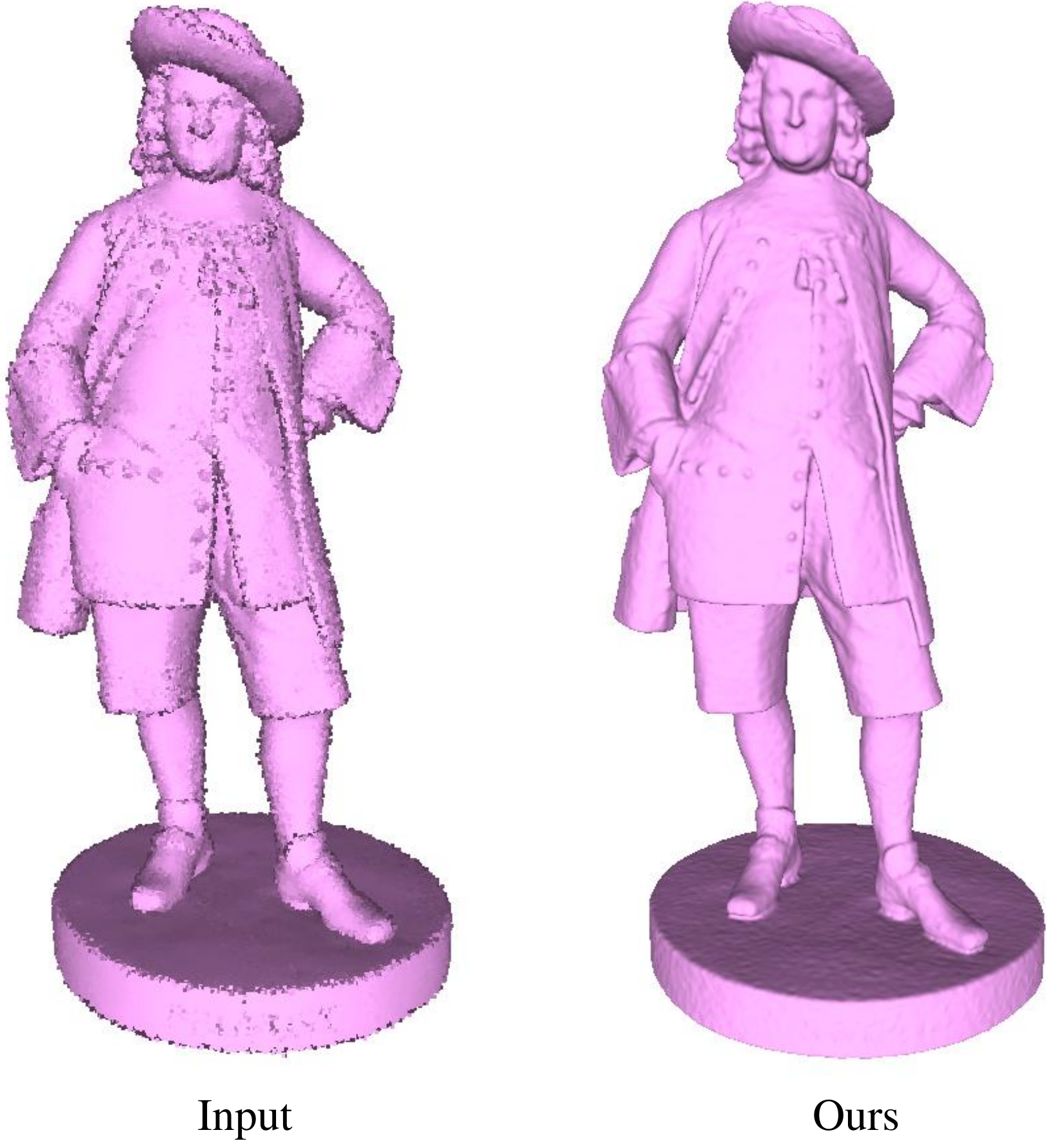}
  \caption{\label{fig:Figure11}
           Our reconstruction of a noisy point cloud with $1$M points.}
\end{figure}

\section{Experiments} \label{experiments}

\subsection{Overview} \label{experiments_overview}

In the experiments, the minimum octree depth $d_{min}$ is set to 6 and the maximum depth $d_{max}$ is set to 8 in the adaptive depth selection strategy of our approach. We perform five alternative updating processes. Specifically, we use $P^{(5)}$ in Equation~\ref{equation4} as the denoised point cloud and $S^{(5)}$ as the reconstructed surface. If $d^{(0)}$ is calculated to be 6 based on the normal variations, then the depths in the subsequent five reconstructions $S^{(1)}$ to $S^{(5)}$ are set to $[6,6,7,7,8]$. If $d^{(0)} = 7$, then the subsequent depths are $[7,7,8,8,8]$. If $d^{(0)} = 8$, all the iterations are carried out within depth $8$. The threshold $d_{sharp}$ in the $\lambda$-projection is set to $8$. Specifically, we only consider the sharp features when $d=8$. 

We choose the following non-data-driven and data-driven approaches as baselines and conduct qualitative and quantitative comparisons with these methods. These approaches belong to different categories.

\textbf{RIMLS}~\cite{2009KernelRegre}: RIMLS applies the robust local kernels to the MLS surfaces. This approach requires consistently oriented normals as inputs. Here, the normals are estimated by PCA~\cite{LIII1901On} and oriented by Hoppe et al.~\cite{1992MST}. The RIMLS implementation of MeshLab~\cite{2008Meshlab} is used to conduct experiments.

\textbf{WLOP}~\cite{2009WLOP}: WLOP is a locally optimal projection method for point cloud consolidation, which generates an evenly distributed particle set of the original point cloud. We use the implementation of WLOP in CGAL~\cite{2009CGAL}.

\textbf{Jet smoothing}~\cite{2005Jets, 2008Jetcpp}: Osculating jet is an efficient polynomial fitting technique to estimate the local surface properties. This technique can be applied to denoise the point cloud through projecting the points onto the local surface shape. We use the ``jet smoothing" implementation of CGAL. 

\textbf{Bilateral smoothing}~\cite{2009CGAL}: Bilateral filter is a useful tool for reducing the noise in point sets. We apply the CGAL implementation of the``bilateral smoothing" filter. It is worth noticing that the CGAL implementation of bilateral smoothing is combined with an edge preserving module based on the philosophy of EAR~\cite{2013EdgeAware}, which is introduced in the official documentation of CGAL. The bilateral smoothing of CGAL also requires normals as inputs. We estimate the normals by PCA and Hoppe et al.~\cite{1992MST} similar to RIMLS.

\textbf{PointCleanNet}~\cite{2020Pointcleannet}: PointCleanNet is a representative supervised learning approach for point cloud filtering. The network contains an outlier detector branch and a denoiser branch both based on PointNet~\cite{2017PointNet}. The trained model provided by the authors are directly applied to conduct the qualitative and quantitative comparisons.

\subsection{Ablation study with iPSR and PGR} \label{iPSR}

Our method performs iPSR~\cite{2022ipsr} in an iterative manner. Therefore, it is necessary to validate that our method cannot be substituted by simply adjusting some parameters in iPSR. The reconstruction results of the iPSR and ours in a noisy horse model are illustrated in Fig.~\ref{fig:Figure4}. In Poisson surface reconstruction, ``point weight'' is an important parameter to control the degree of fitting the sample points. In our method, the point weight is set to $1.0$ during the alternative updating process. If the point weight is reduced to $0.5$, and iPSR is carried out only once, then the reconstructed mesh remains to be noisy. We also show the reconstruction of iPSR in zero point weight. Setting the point weight to zero directly in the iPSR may cause difficulties in algorithm convergence. Accordingly, we manually initialize the normals with a correct orientation. It can be seen that the zero weight iPSR cannot achieve an on par result with our method. Moreover, the ears of the horse are oversmoothened in the zero weight iPSR.

Fig.~\ref{fig:Figure4b} shows the reconstruction of our method and iPSR within different octree depths of a noisy teapot model. In our method, the adaptive depth selection strategy is applied, and the depth $d^{(0)}$ is calculated as $8$. We also show the iPSR reconstructions from octree depth $6$ to $8$. Although reducing the octree depth can make the reconstruction of iPSR smoother, some details are lost due to the low depth setting, for instance, the mouth of the teapot. Therefore, our method can not be substituted by simply adjusting the octree depth of iPSR.

Additionally, we compare our method with another recent unoriented reconstruction approach PGR~\cite{2022pgr}. Similar to iPSR, the point positions are fixed during the optimization in PGR. We set the parameter $k_{w}$ in PGR to 64 and conduct experiments with four values of the parameter $\alpha$: $2.0,4.0,8.0$ and $16.0$. The purpose of adjusting $\alpha$ is to manage the noisy inputs. Fig.~\ref{fig:Figure12} shows the experiment results where we add randomized Gaussian noise with a standard deviation of $1.0 \times 10^{-2}$ in the input. We also colorize the error from the reconstructed mesh to the ground truth. It can be observed that increasing the value of $\alpha$ improves the noise adaptability of PGR and the smoothness of the generated surface. However, it also results in an increase in the distance between the generated surface and the ground truth mesh. On the other hand, our method iteratively updates the point positions, resulting in good visual effects and low distance errors between the generated surface and the ground truth.

\subsection{Comparison with other denoising approaches} \label{compare_denoise}

\begin{table*}[t]
\centering
\caption{Quantitative comparisons on the famous dataset within different noise standard deviations (STD). We report the root mean square distance-to-surface (RMSD) of each method. The RMSD values are multiplied by $10^{3}$.}
\label{table1}
\begin{tabular}{cccccc}
\hline
Method/STD & $0.5 \times 10^{-2}$ & $1.0 \times 10^{-2}$ & $1.5 \times 10^{-2}$ & $2.0 \times 10^{-2}$ & $2.5 \times 10^{-2}$ \\
\hline

RIMLS  & $2.456$ & $3.121$ & $4.378$ & $6.469$ & $9.813$\\

WLOP  & $3.433$ & $3.669$ & $4.032$ & $4.799$ & $6.420$\\

Bilateral-small  & $2.064$ & $2.895$ & $4.298$ & $6.202$ & $8.548$ \\

Bilateral-large  & $3.630$ & $4.061$ & $4.626$ & $5.305$ & $6.447$ \\

Jet-small  & $1.994$ & $2.752$ & $3.837$ & $5.350$ & $7.248$ \\

Jet-large  & $2.807$ & $3.351$ & $4.012$ & $4.892$ & $6.072$ \\

PointCleanNet  & $2.518$  & $3.362$ & $4.267$ & $5.516$ & $7.210$\\

Ours  & $\textbf{1.975}$  & $\textbf{2.245}$ & $\textbf{2.592}$ & $\textbf{3.459}$ & $\textbf{4.253}$ \\
\hline
\end{tabular}
\label{table_MAP}
\end{table*}

\begin{table*}[t]
\centering
\caption{Quantitative comparisons of the reconstructed mesh quality. We show the Chamfer distance (CD) and the normal consistency (NC) of each method. The CD values are multiplied by $10^{3}$. The red color is used to represent the best value amongst all approaches, and the blue is used to denote the second best. The results indicate that our method exhibits well-rounded performance.}
\label{table2}
\resizebox{\linewidth}{!}{
\begin{tabular}{cccccccccccccccccc}
\hline
\multirow{2}{*}{Model} & \multirow{2}{*}{Noise} & \multicolumn{2}{c}{RIMLS} & \multicolumn{2}{c}{WLOP} & \multicolumn{2}{c}{Bilateral-small} & \multicolumn{2}{c}{Bilateral-large} & \multicolumn{2}{c}{Jet-small} & \multicolumn{2}{c}{Jet-large} & \multicolumn{2}{c}{PointCleanNet} & \multicolumn{2}{c}{Ours}\cr
\cmidrule(lr){3-4} \cmidrule(lr){5-6} \cmidrule(lr){7-8} \cmidrule(lr){9-10} \cmidrule(lr){11-12} \cmidrule(lr){13-14} \cmidrule(lr){15-16} \cmidrule(lr){17-18}
 & & CD$\downarrow$ & NC$\uparrow$ & CD$\downarrow$ & NC$\uparrow$ & CD$\downarrow$ & NC$\uparrow$  & CD$\downarrow$ & NC$\uparrow$ & CD$\downarrow$ & NC$\uparrow$  & CD$\downarrow$ & NC$\uparrow$  & CD$\downarrow$ & NC$\uparrow$  & CD$\downarrow$ & NC$\uparrow$ \\
\hline

tortuga & $0.5 \times 10^{-2}$ & $3.98$ & $\textcolor[rgb]{0,0,1}{0.986}$ & $4.33$ & $0.983$ & $4.04$ & $\textcolor[rgb]{0,0,1}{0.986}$  & $5.40$ & $0.984$  & $\textcolor[rgb]{0,0,1}{3.48}$ & $\textcolor[rgb]{0,0,1}{0.986}$  & $3.51$ & $\textcolor[rgb]{0,0,1}{0.986}$  & $\textcolor[rgb]{1,0,0}{3.47}$ & $0.983$ & $\textcolor[rgb]{1,0,0}{3.47}$ & $\textcolor[rgb]{1,0,0}{0.987}$\\

Utah\_teapot & $0.5 \times 10^{-2}$ & $3.76$ & $\textcolor[rgb]{0,0,1}{0.978}$ & $4.52$ & $0.973$ & $3.68$ & $\textcolor[rgb]{1,0,0}{0.980}$  & $4.60$ & $0.974$  & $\textcolor[rgb]{0,0,1}{3.57}$ & $0.975$  & $3.65$ & $0.975$  & $3.61$ & $0.971$ & $\textcolor[rgb]{1,0,0}{3.46}$ & $\textcolor[rgb]{0,0,1}{0.978}$\\

horse & $1.0 \times 10^{-2}$ & $4.81$ & $0.977$ & $6.07$ & $0.973$ & $3.72$ & $0.981$  & $5.07$ & $\textcolor[rgb]{0,0,1}{0.983}$  & $\textcolor[rgb]{0,0,1}{3.56}$ & $0.977$  & $3.72$ & $0.981$  & $\textcolor[rgb]{0,0,1}{3.56}$ & $0.976$ & $\textcolor[rgb]{1,0,0}{3.27}$ & $\textcolor[rgb]{1,0,0}{0.985}$\\

angle & $1.0 \times 10^{-2}$ & $5.79$ & $0.940$ & $7.78$ & $0.908$ & $4.07$ & $\textcolor[rgb]{0,0,1}{0.945}$  & $6.15$ & $0.935$  & $\textcolor[rgb]{0,0,1}{3.90}$ & $0.937$  & $4.42$ & $0.935$  & $4.22$ & $0.924$ & $\textcolor[rgb]{1,0,0}{3.47}$ & $\textcolor[rgb]{1,0,0}{0.948}$\\

Armadillo & $1.5 \times 10^{-2}$ & $9.23$ & $0.932$ & $6.84$ & $0.930$ & $6.27$ & $\textcolor[rgb]{0,0,1}{0.941}$  & $9.40$ & $0.932$  & $5.91$ & $0.935$  & $6.56$ & $0.932$  & $\textcolor[rgb]{0,0,1}{5.87}$ & $0.929$ & $\textcolor[rgb]{1,0,0}{5.31}$ & $\textcolor[rgb]{1,0,0}{0.943}$\\

xyzrgb\_dragon & $1.5 \times 10^{-2}$ & $14.97$ & $0.829$ & $9.75$ & $0.814$ & $\textcolor[rgb]{0,0,1}{5.92}$ & $\textcolor[rgb]{0,0,1}{0.860}$  & $8.72$ & $0.839$  & $5.96$ & $0.852$  & $7.20$ & $0.844$  & $6.59$ & $0.837$ & $\textcolor[rgb]{1,0,0}{5.30}$ & $\textcolor[rgb]{1,0,0}{0.864}$\\

hand & $2.0 \times 10^{-2}$ & $28.94$ & $0.793$ & $12.23$ & $0.875$ & $7.56$ & $0.856$  & $10.75$ & $0.872$  & $\textcolor[rgb]{1,0,0}{6.54}$ & $0.868$  & $8.38$ & $\textcolor[rgb]{1,0,0}{0.880}$  & $8.12$ & $0.860$ & $\textcolor[rgb]{0,0,1}{7.01}$ & $\textcolor[rgb]{0,0,1}{0.879}$\\

serapis & $2.0 \times 10^{-2}$ & $9.27$ & $0.957$ & $6.97$ & $0.956$ & $\textcolor[rgb]{0,0,1}{6.40}$ & $0.957$  & $7.22$ & $\textcolor[rgb]{0,0,1}{0.958}$  & $6.55$ & $0.946$  & $6.77$ & $0.954$  & $6.59$ & $0.950$ & $\textcolor[rgb]{1,0,0}{6.19}$ & $\textcolor[rgb]{1,0,0}{0.964}$\\

Liberty & $2.5 \times 10^{-2}$ & $43.42$ & $0.613$ & $8.81$ & $0.801$ & $12.39$ & $0.713$  & $7.75$ & $0.791$  & $10.30$ & $0.715$  & $7.29$ & $0.778$  & $\textcolor[rgb]{1,0,0}{6.11}$ & $\textcolor[rgb]{0,0,1}{0.805}$ & $\textcolor[rgb]{0,0,1}{6.42}$ & $\textcolor[rgb]{1,0,0}{0.813}$\\

galera & $2.5 \times 10^{-2}$ & $19.64$ & $0.908$ & $7.52$ & $\textcolor[rgb]{0,0,1}{0.937}$ & $7.42$ & $0.923$  & $8.49$ & $0.934$  & $\textcolor[rgb]{0,0,1}{7.09}$ & $0.913$  & $7.44$ & $0.927$  & $7.14$ & $0.927$ & $\textcolor[rgb]{1,0,0}{6.63}$ & $\textcolor[rgb]{1,0,0}{0.941}$\\


\hline
\end{tabular}
}
\label{table_MAP}
\end{table*}

In this section, we compare our method with the other point cloud denoising techniques mentioned in Section~\ref{experiments_overview} to examine the efficacy of our method. We use the \emph{famous} dataset complied by a recent study Points2Surf~\cite{2020P2S}, which contains 22 well-known shapes such as Armadillo, Stanford Bunny and Utah teapot. These shapes are normalized with the maximum axis length to be 1.0, and $10$K to $100$K points are randomly sampled from each shape. Then, the randomized Gaussian noise is added to each sample point of the shape. The standard deviations (STD) are selected within five different levels from $0.5 \times 10^{-2}$ to $2.5 \times 10^{-2}$.  

Parameter settings are typically crucial for non-data-driven denoising approaches. During the experiments, we set the parameter ``filter scale'' for RIMLS to 7. In WLOP, the ``representative particle number'' is set to $95\%$ of the original point set, and the ``neighbor radius'' is set to $0.04$. Jet smoothing has only one parameter ``neighbor size''. ``Jet-small'' is used to represent the parameter setting 48 to the neighbor size. Meanwhile, ``jet-large'' is used to represent the parameter setting 96. Three parameters, namely, ``neighbor size'', ``sharpness angle'' and ``iters'', exist in bilateral smoothing. Increasing the sharpness angle will reduce the sharpness of the results. ``Bilateral-small'' is used to represent the parameter setting with neighbor size 24, sharpness angle 25 and iters 5, whilst ``bilateral-large'' is used to denote neighbour size 48, sharpness angle 50 and iters 5. Only one parameter ``iters'' exists in PointCleanNet, which is set to 5 during the experiments. In our method, the Poisson point weight is set to 1.0 in the datasets with a standard deviation less than $2.0 \times 10^{-2}$, and 0.5 otherwise. Parameters $c$ and $\sigma$ in Equation~\ref{equation8b} are determined as follows. Firstly, the sharpness ratios of all points in a shape is sorted. Then, $c$ is set to the value of the 90\% position of the sorted array, and $\sigma$ is set to $c/2$.

Firstly, we quantitatively compare the quality of the denoised point cloud in terms of the root mean square distance-to-surface (RMSD). If $P$ is the denoised point cloud and $P^{\prime}$ is a densely sampled point cloud of the ground truth mesh. Then, the RMSD value can be calculated as follows:
\begin{equation}
\label{equation_RMSD}
RMSD(P, P^{\prime}) = \sqrt{\frac{1}{N} \sum \limits_{p_i \in P} \min \limits_{p_j \in P^{\prime}} {||p_i - p_j||_{2}^{2}} }.
\end{equation}

The quantitative results are shown in Table~\ref{table1}. The RMSD values are multiplied by $10^3$. The results indicate that our method exhibits high performance in all the five noise scales from $0.5 \times 10^{-2}$ to $2.5 \times 10^{-2}$. The qualitative comparisons are shown in Fig.~\ref{fig:Figure5}. For jet smoothing and bilateral smoothing, we show the better performing from the small parameter setting and the large parameter setting of each shape. We annotate the RMSD value (also multiplied by $10^3$) in the bottom of each point set and colorize the point to surface distance from the denoised point cloud to the ground truth surface. The five examples belong to the dataset of different noise scales. The results show that the generated point clouds of our method perform the lowest error. To make the comparisons more comprehensive, we also conduct an experiment in terms of the mean absolute distance-to-surface (MADS) of the denoised point cloud in Table 3, which is located in the supplementary material. The formulation for calculating the MADS are given as follows:

\begin{equation}
\label{equation_MADS}
MADS(P, P^{\prime}) = \frac{1}{N} \sum \limits_{p_i \in P} \min \limits_{p_j \in P^{\prime}} {||p_i - p_j||_{2}}.
\end{equation}

Furthermore, we compare the quality of the reconstructed mesh amongst different approaches. For WLOP, bilateral smoothing, jet smoothing and PointCleanNet, we use iPSR to generate the reconstructed surfaces from the denoised point clouds provided by their algorithms. The reason is that no consistently oriented normals are required for iPSR. The L1 Chamfer distance (CD) and the normal consistency (NC) are used to measure the mesh quality. Here, L1 CD means to apply the L1 sum toward all sample points, rather than utilizing the L1 distance of point positions. The formula for calculating the CD value is presented as follows:
\begin{small}
\begin{equation}
\label{equation_CD}
CD(P, P^{\prime}) = \frac{1}{|P|} \sum \limits_{p_i \in P} \min \limits_{p_j \in P^{\prime}} {||p_i - p_j||_{2}} 
+ \frac{1}{|P^{\prime}|} \sum \limits_{p_j \in P^{\prime}} \min \limits_{p_i \in P} {||p_j - p_i||_{2}}.
\end{equation}
\end{small}
Where $P$ and $P^{\prime}$ are point clouds uniformly sampled from the reconstructed mesh and the ground truth mesh. Normal consistency (NC) can also be named as the mesh cosine similarity, which calculates the average absolute normal dot product between the sample of the ground truth mesh and the nearest point in the sample of the reconstructed mesh.

Table~\ref{table2} quantitatively compares the mesh quality in terms of CD and NC amongst different approaches. We show the model names, noise scales, CD and NC values in the table. The CD values are also multiplied by $10^{3}$. In each row, the red number represents the best value amongst all methods, and the blue number represents the second best. Our method achieves well-rounded performance. The meshes generated by bilateral smoothing show high NC values. However, the CD values are typically large, indicating that the filtered point positions are far from the ground truth. The results of jet smoothing exhibit low CD values. However, the NC values of the generated surfaces are not that satisfactory. This phenomenon is also shown in Fig. 15, located in the supplementary material, where we present the qualitative results of rows 3,5 and 9 in Table~\ref{table2}. We also colorize the error from the reconstructed mesh to the ground truth mesh with a color bar. The results of our method exhibits low error and high normal consistency. PointCleanNet achieves the lowest CD value in the ``Liberty'' model. But in reality, this shape is in the training set of PointCleanNet. To make the comparisons more comprehensive, we also provide an evaluation based on the F-score of the reconstructed mesh in Table 4, which is also located in the supplementary material. The F-score is calculated as follows: Firstly, we sample a point cloud $P$ from the reconstructed mesh and a point cloud $P^{\prime}$ from the ground truth mesh. For each point $p_i$ in $P$, we measure the distance between $p_i$ and its nearest point $p^{\prime}_{i}$ in $P^{\prime}$. We then measure the proportion of points in $P$ that have a nearest distance below a threshold of $5 \times 10^{-3}$. This proportion is referred as the “precision”. Swapping $P$ and $P^{\prime}$ allows us to assess the “recall”. The F-score, denoted as $F1$, is calculated as:
\begin{equation}
\label{equation_F_score}
F1 = \frac{2 \times precision \times recall}{precision + recall}.
\end{equation}

\subsection{Managing different situations} \label{situations}
In this section, we examine the ability of our method in managing various situations including misalignment, outliers, highly inconsistent sampling densities, noisy CAD-like models and real scanned point clouds. 

\textbf{Misalignment}
Misalignment is a typical noise category, especially for the point clouds obtained with a scanner. Here, we use the misalignment dataset provided by a recent benchmark~\cite{2022PointReconSurvey} to examine the ability of our method to manage this type of structured noise. The point clouds are obtained with the Blensor simulator~\cite{2011Blensor} by adding some perturbations to the camera extrinsics. Fig.~\ref{fig:Figure10} shows the misalignment inputs and the denoised point clouds generated by our method. The results indicate that the misalignment situation is efficiently managed by our method.

\textbf{Outliers and highly inconsistent sampling densities}
Fig.~\ref{fig:Figure6} shows a point cloud including both misalignment artifacts and outliers. In this example, $1$K outliers are randomly sampled within a unit cube and added to a point set with approximately $160$K points. In Poisson surface reconstruction, a few outliers may not have significant influence on the global implicit function. Accordingly, our method performs a certain degree of robustness to outliers, whilst the other traditional approaches encounter some challenges to manage this situation. Jet smoothing and bilateral smoothing can filter out more outliers if a large neighbor size is utilized. For instance, setting the neighbor size of jet smoothing to 1024 can filter out most of the outliers in this case. However, the original shape will also be severely oversmoothed by such a large neighbor scale. Actually, this situation can also be managed by applying iPSR to the denoised point clouds provided by jet and bilateral smoothing, and then projecting the noisy points onto the surface. However, this operation aligns with the underlying philosophy of our method, which further confirms the validity of our concept.

We also conduct experiments to evaluate the performance of our method on point clouds with highly inconsistent sampling densities. In Fig.~\ref{fig:Figure14}, the input point clouds contain randomized Gaussian noise with inconsistent sampling. Notably, the sampling density of one half of the shape is only 1/5 of the other half. The results demonstrate the effectiveness of our method in addressing this challenging issue.

\textbf{Noisy CAD-like inputs}
The CAD-like inputs always include rich sharp features. We have demonstrated in Section~\ref{lambdaprojection} and Fig.~\ref{fig:Figure3} that the $\lambda$-projection method we proposed is helpful for alleviating the oversmoothing phenomenon near the sharp edges. In this section, we mainly focus on the comparisons. We set the parameters $c$ and $\sigma$ in Equation~\ref{equation8b} to be $0.11$ and $0.05$, respectively. Fig.~\ref{fig:Figure7} presents the qualitative comparisons of our method with WLOP and jet smoothing in two CAD-like models. The results illustrate that our method has a fundamental edge reconstruction capability compared with traditional denoising approaches with the help of the $\lambda$-projection.

To enhance the preservation of sharp edges or feature lines, we can incorporate a point resampling process in the future. A recent method RFEPS~\cite{2022RFEPS} specifically addresses this issue in CAD-like input, with a particular focus on edge preservation. In Figure~\ref{fig:Figure13}, we provide qualitative comparisons between our method and RFEPS. We utilize the source code provided by the authors and keep the default parameters. While RFEPS demonstrates favorable edge preservation, its applicability to general shapes may be limited. Furthermore, we observe that the edge point detection and sampling process of RFEPS could potentially be integrated into our alternative updating process in the future. Enhancing the preservation of sharp features is one of our ongoing research directions.

\textbf{Real scanned data} 
In Fig.~\ref{fig:Figure9}, we examine the ability of our approach in handling the real scanned point clouds and compare our method with iPSR (run only once) and jet smoothing. The data are provided by~\cite{2022PointReconSurvey}. The surfaces of jet smoothing are generated by feeding the denoised point clouds to iPSR. Our method achieves decent performance. Jet smoothing even aggravates the noise near the thin structures, resulting in the failure of the iPSR to converge to the correct surface for the bowl model.

\section{Conclusion} \label{conclusion}
In this work, we propose an alternative denoising and reconstructing approach for unoriented point sets and performs iPSR in an iterative manner. An adaptive depth selection strategy is proposed to ensure that the reconstruction is carried out within an appropriate octree depth of iPSR. Moreover, we present a $\lambda$-projection method to handle the challenge of oversmoothing near the sharp edges during the iterative process. The experimental results show that our method exhibits high performance in point cloud denoising and surface reconstruction tasks and manages various situations.

The main drawback of our method is the high time consumption compared with traditional denoising techniques. Our method performs iPSR several times and acts as an outer loop of iPSR. iPSR itself is an iterative method for the screened Poisson surface reconstruction. Accordingly, our method requires about $8.5$min on AMD Ryzen 5 5600H CPU @ 3.3GHz to denoise and reconstruct the bowl model of Fig.~\ref{fig:Figure9} with approximately $0.2$M points. However, one time iPSR reconstruction is also required for other denoising approaches to generate a reliable surface when no consistently oriented normals are provided. For instance, combining jet smoothing and iPSR also takes about $105$s for this model. Our method is faster than the learning-based PointCleanNet, which requires $27$min in the RTX 2080Ti GPU with 5 iters. Furthermore, the reconstruction complexity of screened Poisson surface reconstruction is a linear function of the point number due to the application of the conforming cascade Poisson solver. Therefore, the time complexity of our method is not large relative to the point number. Fig.~\ref{fig:Figure11} demonstrates our reconstruction of a noisy point cloud with $1$M points. The denoising process is carried out within octree depth 10. Our method can manage this example in about one hour.

In addition to high time consumption, our current method is limited to a fixed number of iterations in the outer loop. In the future, we can enhance the adaptiveness of our method by terminating the denoising process dynamically with an evaluation mechanism. To further improve the sharp feature preservation, we can integrate the feature-line detection and resampling techniques to our method.

\bibliographystyle{cag-num-names}
\bibliography{refs}

\newpage
\includepdf[pages=-]{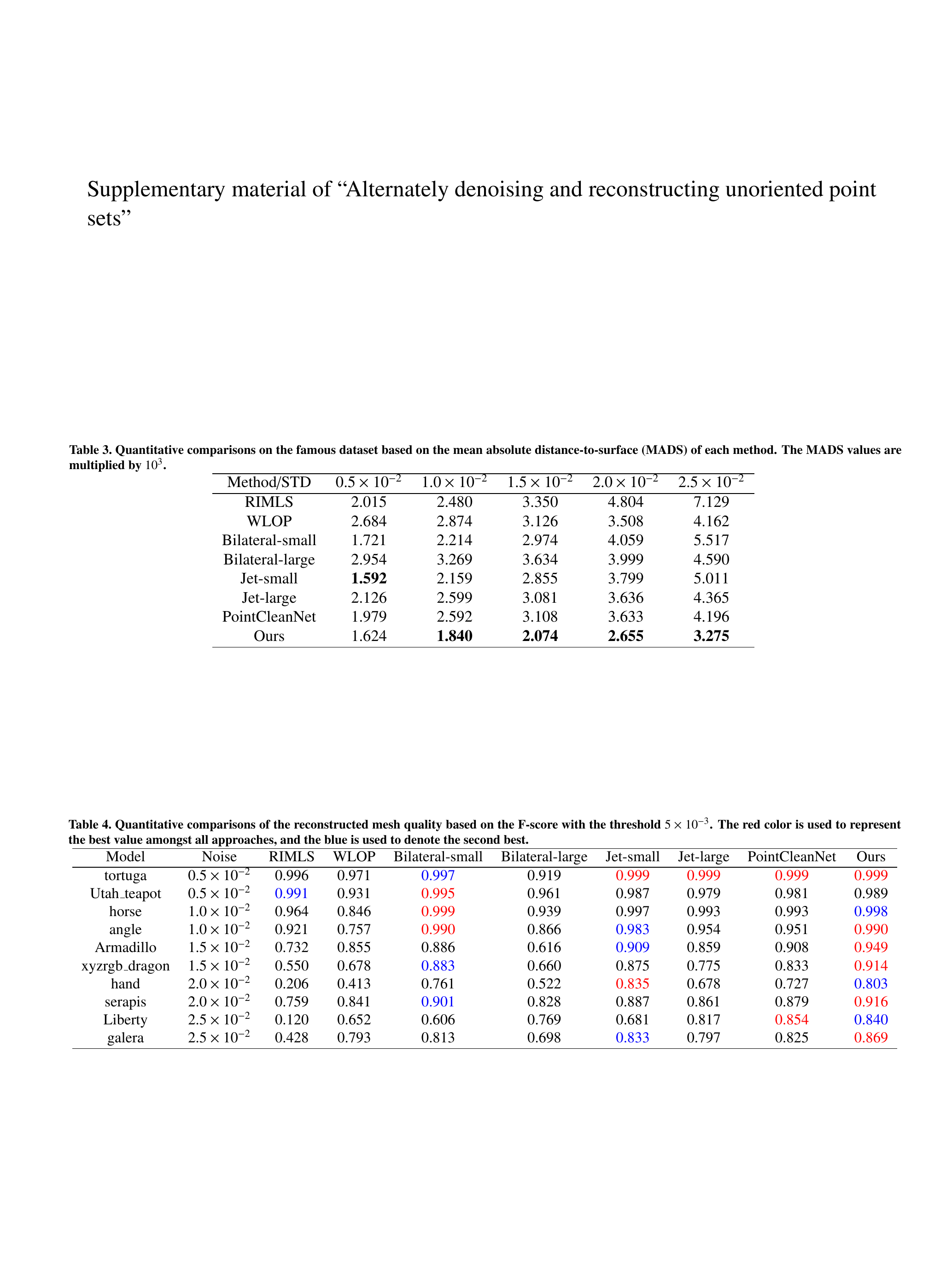} 

\end{document}